\documentstyle[12pt,fleqn]{article}
\def\be{\begin{equation}}
\def\ee{\end{equation}}
\def\ba{\begin{eqnarray}}
\def\ea{\end{eqnarray}}

\def\fun#1#2{\lower3.6pt\vbox{\baselineskip0pt\lineskip.9pt

\ialign{$\mathsurround=0pt#1\hfill##\hfil$\crcr#2\crcr\sim\crcr}}}
\begin{document}

\setbox16= \hbox{der \kern -17.2pt \raise %
 3.8pt \hbox{-} \vrule width -2.9pt depth 0pt }
\def\dj{\copy16}

\begin{titlepage}
\null\vspace{-62pt}
\begin{flushright}LBL-34592;\ UCB-PTH-93/24\\
      CMU--HEP93--11;\ DOE--ER/40682--36\\
      August 1993

\end{flushright}
\vspace{0.3in}
\centerline{{\large \bf Approximate Flavor Symmetries }}
\centerline{{\large \bf in the Lepton Sector}}
\vspace{0.4in}
\centerline{
Andrija Ra\v{s}in$^{(a)}$$^{*}$\ \ and \ \ Jo\~{a}o P. Silva$^{(b)}$}
\vspace{0.3in}
\centerline{{\it $^{(a)}$Department of Physics,
University of California, Berkeley }}
\centerline{{\it and Lawrence Berkeley Laboratory, Berkeley, CA~~94720}}
\vspace{.2in}
\centerline{{\it $^{(b)}$Physics Department, Carnegie Mellon University,
Pittsburgh, PA~~ 15213}}
\vspace{.5in}
\baselineskip=24pt

\centerline{\bf Abstract}
\begin{quotation}
Approximate flavor symmetries in the quark sector have been used
as a handle on physics beyond the Standard Model.
Due to the great interest in neutrino masses and mixings and the
wealth of existing and proposed neutrino
experiments it is important to extend
this analysis to the leptonic sector.
We show that in the see-saw mechanism, the neutrino masses and mixing
angles do not depend on the details of the right-handed
neutrino flavor symmetry breaking, and are related by a simple formula.
We propose several ans\"{a}tze which relate different flavor
symmetry breaking parameters and find that the MSW solution to
the solar neutrino problem is always easily fit.
Further, the $\nu_\mu - \nu_\tau$ oscillation is unlikely
to solve the atmospheric neutrino problem and, if we fix the
neutrino mass scale by the MSW solution, the neutrino masses
are found to be too small to close the Universe.

PACS number(s): 11.30.Hv, 12.15.Ff, 13.90.+i
\end{quotation}
\end{titlepage}


\newpage

\begin{center}
{\bf Disclaimer}
\end{center}

\vskip .2in

\begin{scriptsize}
\begin{quotation}
This document was prepared as an account of work sponsored by the United
States Government.  Neither the United States Government nor any agency
thereof, nor The Regents of the University of California, nor any of their
employees, makes any warranty, express or implied, or assumes any legal
liability or responsibility for the accuracy, completeness, or usefulness
of any information, apparatus, product, or process disclosed, or represents
that its use would not infringe privately owned rights.  Reference herein
to any specific commercial products process, or service by its trade name,
trademark, manufacturer, or otherwise, does not necessarily constitute or
imply its endorsement, recommendation, or favoring by the United States
Government or any agency thereof, or The Regents of the University of
California.  The views and opinions of authors expressed herein do not
necessarily state or reflect those of the United States Government or any
agency thereof of The Regents of the University of California and shall
not be used for advertising or product endorsement purposes.
\end{quotation}
\end{scriptsize}

\vskip 2in

\begin{center}
\begin{small}
{\it Lawrence Berkeley Laboratory is an equal opportunity employer.}
\end{small}
\end{center}


\newpage
\setcounter{page}{1}

\baselineskip=24pt

\section{Introduction}

The smallness
of the Yukawa couplings has been related to
the approximate flavor symmetries (\cite{frog79}, \cite{anta92}).
The Yukawa couplings can be understood as being
naturally small \cite{thoo80},
since by putting them equal to zero new
global symmetries appear in the theory.
These are the flavor
symmetries under which each of the fermions
transforms separately but scalars do not.
In the low energy theory these symmetries are broken
by different amounts as is evident from the nonzero
masses of fermions. The lack of knowledge of
the exact mechanism by which the symmetries were
broken was parametrized by a set of small numbers
(denoted by $\epsilon$).
Each Yukawa coupling is then given approximately as
the product of small symmetry breaking parameters
$\epsilon$ for all flavor symmetries broken
by that coupling. So for example the coupling of a scalar doublet
$H$ to the ith generation of left-handed doublet quarks
$Q_i$  and the jth generation of right-handed up quark $U_j$ is
given by
\begin{equation}
\lambda^U_{ij} \approx \epsilon_{Q_i} \epsilon_{U_j} .
\end{equation}

Whereas this gives the right order of magnitude for the
couplings, the exact relation of couplings to the $\epsilon$s
may be off by a factor of 2 or 3.
This is because the underlying theory (possibly a GUT) by which
the flavor symmetries
are broken is unknown. Therefore all estimates of
the possible
new flavor changing interactions should be taken to have at least
the same
uncertainty.

The flavor symmetry breaking parameters $\epsilon$ can be estimated
in several ways. They can
be postulated by ans\"{a}tze \cite{anta92}
which are consistent with the known values of fermion masses and mixings.
Alternatively, one may
use the experimental results to constrain
the $\epsilon$s. This was done in the quark sector \cite{hall93}
using the known values of quark masses and Kobayashi-Maskawa
mixing matrix elements.

The important result so obtained was that
flavor changing scalar interactions
are possible even for the masses of the new scalars
as low as a few hundred GeV to 1 TeV, and numerous
estimates for different flavor changing interactions were
given \cite{anta92,hall93}.
Hall and Weinberg \cite{hall93} also noticed that allowing
for complex Yukawa couplings in this scheme, the
observed smallness of CP violation constrains the
phases to be small (i.e of the order of $10^{-3}$ ).

In the lepton sector, the masses are small indicating that the approximate
flavor symmetries are preserved to a high degree of acuracy.
Therefore lepton flavor changing interactions would be extremely
hard to test \cite{anta92}. Further, since
the neutrino masses and mixing angles have not been
measured yet, one cannot estimate the corresponding
$\epsilon$s without additional assumptions.
Nevertheless we find it important to address the question
of approximate flavor symmetries in the lepton sector, because
many experiments which aim to measure or set better limits
on neutrino masses and mixings are under way or being planned
for the near future. Indeed, in this paper we find that statements
can be made about the lepton sector, regardless of additional
assumptions about the $\epsilon$s.
For example, while the MSW
solution to the solar neutrino problem can be easily fit, the predicted
neutrino masses are unlikely to close the Universe.

This paper is organised as follows.
In section 2 we express the lepton sector Yukawa couplings
in terms of flavor symmetry breaking parameters. Here we study
two cases of neutrino masses.
In the see-saw mechanism \cite{gell79} case we show
that the neutrino masses and mixings are independent of the
right-handed neutrino flavor symmetry breaking mechanism.
We also include for completness the case of
Dirac neutrino masses only (as in the case of charged fermions).
In section 3 we provide several
ans\"{a}tze for the lepton flavor symmetry
breaking parameters and list their predictions in terms
of ratios of neutrino masses, and mixings. We study their
relevance to the solar neutrino problem, atmospheric neutrino
problem, closure of the Universe, etc. General features
are noted which are independent of the particular ansatz used.

\section{Approximate Flavor Symmetries in the Lepton Sector}

By adding the right-handed neutrinos $N_i$,
$i=1,2,3$ to the particle content of the
standard model we can allow for Dirac
type masses. Under the action of approximate
flavor symmetries, whenever an $N_i$ enters
a Yukawa interaction, the corresponding
coupling must contain the symmetry
breaking parameter $\epsilon_{N_i}$.

A natural way
to justify the smallness of neutrino masses is
to use the {\it see-saw mechanism} in which the smallness
of the left-handed neutrino masses is explained by the
new scale of heavy right-handed neutrinos.
The mass matrices will have the following
structure
\begin{equation}
m_{N_{Dij}} \approx \epsilon_{L_i} \epsilon_{N_j} v_{SM} ,
\end{equation}
\begin{equation}
m_{N_{Mij}} \approx \epsilon_{N_i} \epsilon_{N_j} v_{Big} ,
\end{equation}
\begin{equation}
m_{E_{ij}} \approx \epsilon_{L_i} \epsilon_{E_j} v_{SM} ,
\end{equation}
where $m_{N_D}$ and $m_E$ are the neutrino and charged
lepton Dirac mass matrices,
$m_{N_M}$ is the right-handed neutrino Majorana mass matrix,
$v_{SM}=174 \ GeV$
and $v_{Big}$ is the new large mass scale.
The generation indices $i$ and $j$ run from 1 to 3.
In the following we assume a hierarchy in the $\epsilon$s
(i.e. $\epsilon_{L_1} << \epsilon_{L_2} << \epsilon_{L_3}$, etc.)
as suggested by the hierarchy of quark and charged lepton masses.
Then the diagonalization of the neutrino mass matrix will give
a heavy sector with masses
$m_{N_{Hi}} \approx \epsilon^2_{N_i} v_{Big}$ and a very light
sector with mass matrix
\begin{equation}
m_{N_{Lij}} \approx ( m_{N_D} m^{-1}_{N_M} m^T_{N_D} )_{ij}
\approx \epsilon_{L_i} \epsilon_{L_j} \frac {v^2_{SM}} {v_{Big}} ,
\end{equation}
where the number
$Tr(\epsilon_N (\epsilon_N \epsilon_N)^{-1} \epsilon_N)$ is assumed to be
of order one. We have the expected result: the heavy right-handed
neutrino decouples from the theory leaving behind a very
light left-handed neutrino. The masses and mixing angles are
independent of the right-handed symmetry breaking parameters
$\epsilon_{N_i}$ :
\begin{eqnarray}
m^N_i
& \approx &
\epsilon^2_{L_i} \frac {v^2_{SM}} {v_{Big}}\ ,
\nonumber\\
m^E_i
& \approx &
\epsilon_{L_i} \epsilon_{E_i} v_{SM}
\ \ \ (no \  sum \  on \  i)\ ,
\nonumber\\
V_{ij}
& \approx &
\frac {\epsilon_{L_i}} {\epsilon_{L_j}}\ \ \  (i<j)\ .
\label{eq:majorana}
\end{eqnarray}
Therefore,
besides the unknown scale $v_{Big}$, only two sets of
$\epsilon$s are needed : $\epsilon_{L_i}$ and $\epsilon_{E_i}$.
In fact, the neutrino masses and mixings depend only on $\epsilon_{L_i}$
and they are approximately related through
\begin{equation}
V_{ij} \approx \sqrt{ \frac {m^N_i} {m^N_j} } .
\label{eq:relation}
\end{equation}
%
Equation (7) reduces the number of parameters needed to describe neutrino
masses and mixings by three; for example, given two mixing angles and
one neutrino mass, we can predict the third mixing angle and the other
two neutrino masses.
These results are extremely general.
They follow simply from the factorization of the Dirac masses,
regardless of the specific form of $m^{-1}_{N_M}$,
which only contributes to set the scale.

For completeness, we note that in the
case of {\it Dirac masses only},
the neutrinos and the charged leptons
mass matrices become
\begin{equation}
m_{N_{Dij}} \approx \epsilon_{L_i} \epsilon_{N_j} v_{SM} ,
\end{equation}
\begin{equation}
m_{E_{ij}} \approx \epsilon_{L_i} \epsilon_{E_j} v_{SM} .
\end{equation}
Their diagonalization yields
\begin{eqnarray}
m^N_i
& \approx &
\epsilon_{L_i} \epsilon_{N_i} v_{SM}
\ \ \ (no \  sum \  on \  i)\ ,
\nonumber\\
m^E_i
& \approx &
\epsilon_{L_i} \epsilon_{E_i} v_{SM}
\ \ \ (no \  sum \  on \  i)\ ,
\label{eq:dirac1}
\end{eqnarray}
whereas the lepton mixing matrix is approximately diagonal
with off diagonal elements of the order of
\begin{equation}
V_{ij} \approx \frac {\epsilon_{L_i}} {\epsilon_{L_j}}\ \ \  (i<j)\ .
\label{eq:dirac2}
\end{equation}
Therefore in this case we need three sets of $\epsilon$s to explain
masses and mixings (cf. Eqns.(\ref{eq:dirac1})-(\ref{eq:dirac2}) ):
$\epsilon_{L_i}$, $\epsilon_{N_i}$ and $\epsilon_{E_i}$.

In addition to having one more set of (unknown) $\epsilon$s
than the {\it see-saw} case, the {\it Dirac only} case has
no natural explanation for the
very low scale associated with the neutrino masses, i.e.
that the flavor symmetries of right-handed neutrinos
are much better preserved than for other fermions.
It turns out that one typically gets
the $\epsilon_{N_i}$ of order $10^{-12}$ or so, compared to
quark and charged lepton $\epsilon$s which are typically
$10^{-3}$ and larger.
Therefore, in the following we will mostly concentrate on the
{\it see-saw} case.

\section{Ans\"{a}tze and Predictions}

While it was possible to determine the flavor symmetry breaking
parameters in the quark sector from quark masses and mixings
\cite{hall93}, in the lepton sector the situation is much more
difficult. At this time direct laboratory experiments provide
only upper limits on neutrino masses and mixings, although some
indirect sources like solar neutrinos or cosmology point to some
specific allowed ranges.
Therefore to  estimate the sizes of lepton flavor symmetry breaking
parameters $\epsilon$s additional assumptions are needed.

Our strategy is as follows: we list several plausible or GUT motivated
ans\"{a}tze
which relate $\epsilon$s of different fields.
This will enable us
to estimate ratios of neutrino masses and mixings.
If the mixings are consistent with the
allowed range for the MSW solution
\cite{mikh80} of the solar neutrino problem,
we take this as a hint
to fix the mass scale and predict all neutrino masses. We then
look at further predictions.
As is the case with any calculation based on these approximate flavor
symmetries ( \cite{frog79}, \cite{anta92},
\cite{hall93} ), the factors of two or three may contribute
coherently factors of an order of magnitude or so.
In addition, the numerical results depend on the specific ansatz.
What we seek are the general features of those results
rather than the detailed numerical results themselves.

{\bf I}. As our first ansatz we assume that $\epsilon_{L_i}=\epsilon_{E_i}$
\cite{anta92}. This ansatz can be justified as follows \cite{anta93}.
Assume that in the lepton sector the only combination
of symmetry which is broken is the axial
flavor symmetry. This means that, instead of breaking
separately left(L) and right(R) flavor symmetries,
only the combination L+R is broken. Therefore
we need only one set of $\epsilon$s,
which are then determined from
$\epsilon_{L_i} \approx \sqrt{\frac{m^E_i}{v_{SM}}}$.
The $\nu_e - \nu_\mu$ mixing found in this way
is consistent with the
small mixing angle region \cite{snp} of the MSW explanation
for the Solar Neutrino Problem (SNP).
The mixing angles for this and the other ans\"{a}tze
can be found in Table~1.
We checked
that for these mixing angles the third flavor does in fact decouple
(see \cite{pant93}).
If this is indeed the solution
for that problem, the mass of the muon neutrino must be around
$3 \times 10^{-3}eV$.
This then sets the scale for the
new physics and the
other neutrino masses at
\ba
v_{Big} & \approx &
\frac{\epsilon_{L_2}^2 v_{SM}^2}{m_{\nu_\mu}}
\approx 10^{13} GeV\ ,\\
m_{\nu_e} & \approx &
\frac{\epsilon_{L_1}^2}{\epsilon_{L_2}^2} m_{\nu_\mu}
\approx 10^{-5} eV\ ,
\nonumber\\
m_{\nu_\tau} & \approx &
\frac{\epsilon_{L_3}^2}{\epsilon_{L_2}^2} m_{\nu_\mu}
\approx 10^{-2} eV\ .
\label{eq:ahr}
\ea
Taking into account the excluded regions due to the IMB experiment
\cite{IMB},
the predicted muon to tau neutrino mixing ($\Delta m^2 \approx
10^{-3}\  eV^2$, $\sin^2(2 \theta_{\mu \tau}) \approx 0.2$)
is around a factor of five away from the parameters
consistent with a $\nu_\mu - \nu_\tau$ oscilation explanation
of the Atmospheric Neutrino Problem (ANP) \cite{anp},
and still two orders of
magnitude away from the laboratory limits for this mixing.
Finally, the smallness of all three neutrino masses in this
ansatz suggests that
neutrinos cannot be responsible for closing the Universe.

{\bf II}. Another interesting ansatz is suggested by the fact that in the
quark sector $\epsilon_{Q_i} \approx \epsilon_{U_i} \ \ i=1,2,3$;
as found by Hall and Weinberg \cite{hall93}.
Inspired by an $SU(5)$ type of unification
we are led to look at an ansatz in which,
\be
\epsilon_{L_i} \propto \epsilon_{D_i}\ ,
\hspace{8mm}
\epsilon_{E_i} \propto \ \epsilon_{Q_i} \approx \epsilon_{U_i}\ .
\ee
In this ansatz we {\it predict} (using the
numerical values of $\epsilon_Q$,
$\epsilon_U$ and $\epsilon_D$ from \cite{hall93} )
the ratios of charged lepton masses
to be within factors of three of the measured values.
We consider this an interesting result.
Further, the mixing angles are
consistent with the three flavour mixing explanations of the
SNP \cite{pant93} for squared masses of $\nu_\mu$ and $\nu_\tau$
of order $10^{-4}\ eV^2$.
Therefore,
the $\nu_\mu - \nu_\tau$
oscilation explanation of the ANP is unlikely. In addition this mass
scale cannot be tested in the laboratory nor provide an explanation
for Dark Matter.

{\bf III}. Finally, one might look for inspiration in the breaking of $SO(10)$
into
$SU(4) \otimes SU(2)_L \otimes SU(2)_R$. We know that
at the renormalization scale
of $1 GeV$ we have been working at,
the $SU(2)_R$ symmetry must be badly broken since $m_t >> m_b$ and
$m_e >> m_{\nu_e}$.
Further, assuming the ansatz,
$\epsilon_{L_i} \propto \epsilon_{Q_i}$,
$\epsilon_{E_i} \propto \epsilon_{D_i}$,
and
$\epsilon_{N_i} \propto \epsilon_{U_i}$
would lead to $m_e / m_\mu \approx 4.8 \times 10^{-2}$,
which is wrong by an order of magnitude.
Assuming that the $SU(4)$ might still provide some
useful information for the $SU(2)_L$ singlets we look at the ansatz,
\be
\epsilon_{E_i} \propto \epsilon_{D_i}\ ,
\hspace{8mm}
\epsilon_{N_i} \propto \epsilon_{U_i}\ .
\ee
In this ansatz we again predict a
$\nu_e  - \nu_\mu$ mixing angle consistent
with the small angle MSW solution to the SNP. Again, assuming that this
is indeed the solution for that problem fixes the see-saw
neutrino masses at
\ba
m_{\nu_\mu} &\approx& 10^{-3} eV\ ,
\nonumber\\
m_{\nu_e} &\approx& (\frac{\epsilon_{L_1}}{\epsilon_{L_2}})^2 m_{\nu_\mu}
\approx 10^{-6} eV\ ,
\nonumber\\
m_{\nu_\tau} &\approx& (\frac{\epsilon_{L_3}}{\epsilon_{L_2}})^2
m_{\nu_\mu}
\approx 0.5 eV\ .
\ea
We again find it unlikely that the values obtained can
close the Universe or solve the ANP.

\vspace{0.5in}

In conclusion, we extended the concept of approximate flavor symmetries
to the lepton sector. In particular we considered the see-saw mechanism
as a source of the neutrino masses and
showed that the predictions do not depend
on the neutrino flavor symmetry breaking parameters.
This yelds a simple relation (cf. eq. (\ref{eq:relation}))
between neutrino masses
and mixing angles which reduces the number of parameters needed to describe
the lepton sector.
The lack of information on the neutrino masses and mixing angles
led us to propose several ans\"{a}tze.
These exibit the following common features.
They are consistent with the MSW solution of the SNP.
The ANP is unlikely to be explained through $\nu_\mu - \nu_\tau$
oscilations and the scale of neutrino masses is too small
to close the Universe.

\vspace{18pt}
\centerline{\bf Acknowledgements}
We would like to thank Aram Antaramian,
Lawrence Hall, Stuart Raby and Lincoln Wolfenstein for useful
discussions.
We warmly thank the organizers of the
TASI 93 Summer School in Boulder, Colorado where
part of this work has been done.
The work of J.\ P.\ S.\ was supported in part by DOE grant
\# DE-FG02-91ER40682 and by the Portuguese JNICT under
CI\^{E}NCIA grant \# BD/374/90-RM.
\vspace{36pt}

\( ^{*} \) On leave of absence from the
Ru\dj \hspace{.1in} Bo\v{s}kovi\'{c} Institute, Zagreb, Croatia.

\newpage

\centerline{Table Captions}
\vspace{0.4in}

Table~1: Neutrino mixing angle predictions in the three ans\"{a}tze
introduced. As noted in the text, these results are meant as
estimates rather than precise calculations.

\newpage

\begin{table}
\centering
\begin{tabular}{|c|c|c|c|}
\hline
ansatz                                                  &
$\sin^2(2 \theta_{e \mu})$                              &
$\sin^2(2 \theta_{e \tau})$                             &
$\sin^2(2 \theta_{\mu \tau})$                           \\
\hline
{\bf I}                                                 &
$2 \times 10^{-2}$                                      &
$10^{-3}$                                               &
$0.2$                                                  \\
{\bf II}                                                &
$0.2$                                                   &
$0.1$                                                   &
$0.8$                                                  \\
{\bf III}                                               &
$2 \times 10^{-3}$                                      &
$8 \times 10^{-6}$                                      &
$2 \times 10^{-2}$                                     \\
\hline
\end{tabular}
\end{table}

\end{document}